\documentclass[onecollarge,natbib]{svjour2}
\bibpunct{[}{]}{,}{n}{}{,} % to get "[numbered]" references from natbib

\smartqed  % flush right qed marks, e.g. at end of proof

\usepackage{graphicx}
\usepackage{mathptmx}      
\usepackage{latexsym}
\journalname{Few-Body Systems} 

\newcommand{\eq}{\begin{eqnarray}}
\newcommand{\en}{\end{eqnarray}}

\begin{document}

\title{Nucleon Generalized Parton Distributions and Holographic Models 
\thanks{This work was supported by FONDECYT (Chile) under Grant No. 1100287. and Grant No. 3100028; by Federal Targeted Program "Scientific
and scientific-pedagogical personnel of innovative Russia" Contract No. 02.740.11.0238.} 
\thanks{Presented by Alfredo Vega at LIGHTCONE 2011, 23 - 27 May, 
2011, Dallas.}
}
\titlerunning{Nucleon Generalized Parton Distributions and Holographic Models} 

\author{Alfredo~Vega$^1$,~Ivan~Schmidt$^1$,~Thomas~Gutsche$^2$,~Valery~E.~Lyubovitskij$^{2,\ast}$\thanks{$\ast$ 
On leave of absence from Department of Physics, Tomsk State University, 634050 Tomsk, Russia}}

\authorrunning{Alfredo Vega, Ivan Schmidt, Thomas Gutsche, Valery E. Lyubovitskij} 
   
\institute{$^1$ Departamento de F\'\i sica y Centro Cient\'\i fico 
                Tecnol\'ogico de Valpara\'\i so (CCTVal), 
                Universidad T\'ecnica Federico Santa Mar\'\i a, 
                Casilla 110-V, Valpara\'\i so, Chile
                \and  \\ 
           $^2$ Institut f\"ur Theoretische Physik,
                Universit\"at T\"ubingen, 
                Kepler Center for Astro and Particle Physics, \\
                Auf der Morgenstelle 14, D-72076 T\"ubingen, Germany
}

\date{Received: date / Accepted: date}

\maketitle

\begin{abstract} 
Using ideas from Light Front Holography, we discuss the calculation of the nucleon helicity-independent generalized parton distributions of quarks in the zero skewness case.

\end{abstract}

\section{Introduction}

Due to their nonperturbative nature the GPDs cannot be directly calculated from Quantum Chromodynamics (QCD). There are essentially three ways to access the GPDs (for reviews see e.g.~\cite{Goeke:2001tz,Ji:2004gf}): extraction from the experimental measurement of hard processes, a direct 
calculation in the context of lattice QCD, and different phenomenological models and methods. The last procedure is based on a parametrization of 
the quark wave functions/GPDs using constraints imposed by sum rules~\cite{Ji:1996nm,Radyushkin:1997ki}, which relate the parton distributions to nucleon electromagnetic form factors (some examples of this procedure can be found e.g. in~\cite{Diehl:2004cx,Guidal:2004nd,Selyugin:2009ic}). On the other hand, such sum rules can also be used in the other direction -- GPDs are extracted by calculating nucleon electromagnetic form factors in some approach. 

Following the last idea, here we show how to extract the quark GPDs of the nucleon in the framework of a holographical soft-wall model~\cite{Vega:2010ns}. In particular, we use the results of Abidin and Carlson for nucleon form factors~\cite{Abidin:2009hr}, in order to extract the GPDs using the light-front mapping -- the key ingredient of light-front holography~(LFH). This is an approach based 
on the correspondence of string theory in Anti-de Sitter~(AdS) space and conformal field theory~(CFT) in physical space time~\cite{Maldacena:1997re}. LFH is further based on a mapping of the string modes in the AdS fifth dimension to the hadron light-front wave functions in physical space-time, as suggested and developed by Brodsky and de T\'eramond~\cite{Brodsky:2003px,Brodsky:2006uqa,Brodsky:2007hb,Brodsky:2008pf} and extended in~\cite{Vega:2009zb,Vega:2010yq,Branz:2010ub}. We discuss how LFH can be used to get the nucleon GPDs in the context of the soft-wall model~\cite{Vega:2010ns}. Notice that our approach is different to discussed in ~\cite{Nishio:2011xa}.

In this talk we perform a matching of the nucleon electromagnetic form factors considering two approaches: we use sum rules derived in QCD~\cite{Ji:1996nm,Radyushkin:1997ki}, which contain GPDs for valence quarks, and we consider an expression obtained in the AdS/QCD soft-wall model~\cite{Abidin:2009hr}. As result of the matching we obtain expressions for the nonforward parton densities $H_{v}^{q}(x,t) = H^q(x,0,t) + H^q(-x,0,t)$ and $E_{v}^{q}(x,t) = E^q(x,0,t) + E^q(-x,0,t)$ -- flavor combinations of the GPDs (or valence GPDs), 
using information coming from the AdS side. The procedure discussed here is similar to the one used in LFH,  which allows to obtain a light front mesonic wave function related to the AdS modes associated with mesons~\cite{Brodsky:2003px,Brodsky:2006uqa,Brodsky:2007hb,Brodsky:2008pf}.  
 
\section{GPDs in AdS/QCD} 

The nucleon electromagnetic form factors $F_1^N$ and $F_2^N$ ($N=p, n$ correspond to proton and neutron) are conventionally defined by the matrix element of the electromagnetic current as 
\eq 
\label{CorrienteEM}
\langle p' | J^{\mu}(0) | p \rangle = 
\bar{u}(p') [ \gamma^{\mu} F_{1}^N(t) 
+ \frac{i \sigma^{\mu \nu}}{2 m_N}  \, q_\nu 
F_{2}^N(t)] u(p),
\en 
where $q = p' - p$ is the momentum transfer; $m_N$ is the nucleon mass; $F_1^N$ and $F_2^N$ are the Dirac and Pauli form factors, which are normalized to electric charge $e_N$ and anomalous magnetic moment $k_N$ of the corresponding nucleon: $F_1^N(0)=e_N$ and $F_2^N(0)=k_N$.  

The sum rules relating the electromagnetic form factors and the GPDs read as~\cite{Ji:1996nm,Radyushkin:1997ki}  
\eq 
F_{1}^{p}(t) &=& \int_{0}^{1} dx \, \biggl( \frac{2}{3}H_{v}^{u}(x,t) 
                                  - \frac{1}{3}H_{v}^{d}(x,t)\biggr)\,, 
\nonumber\\
F_{1}^{n}(t) &=& \int_{0}^{1} dx \, \biggl( \frac{2}{3}H_{v}^{d}(x,t) 
                                  - \frac{1}{3}H_{v}^{u}(x,t)\biggr)\,, 
\nonumber\\                                    
F_{2}^{p}(t) &=& \int_{0}^{1} dx \, \biggl( \frac{2}{3}E_{v}^{u} (x,t) 
                                  - \frac{1}{3}E_{v}^{d} (x,t)\biggr)\,,
\nonumber\\ 
F_{2}^{n}(t) &=& \int_{0}^{1} dx \, \biggl( \frac{2}{3}E_{v}^{d} (x,t) 
\nonumber                         - \frac{1}{3}E_{v}^{u} (x,t)\biggr)\,. 
\en 
Here we restrict our analysis to the contribution of the $u$ and $d$ quarks and antiquarks, while the presence of the heavier strange and charm quark constituents is not considered.

For other side Abidin and Carlson~\cite{Abidin:2009hr} are calculated the nucleon form factors using an AdS/QCD model. It is based on soft-wall breaking of conformal invariance by introducing the quadratic dilaton field $\Phi (z) = \kappa^{2} z^{2}$ in the action (in the overall exponential and in the mass term)~\cite{Abidin:2009hr}. Such a procedure leads to Regge-like mass spectra in the baryonic sector, and it is similar to use a z dependent mass as in \cite{Vega:2008te}. One should stress that introduction of the dilaton field in both approaches is based on the idea of getting the simplest analytical solution of the equations of motion of the string mode. The AdS metric is specified as 
\begin{equation}
\label{Metrica}
 ds^{2} = g_{MN} dx^M dx^N = 
\frac{1}{z^{2}} (\eta_{\mu \nu} dx^{\mu} dx^{\nu} - dz^{2}),
\end{equation}
where $\mu, \nu = 0, 1, 2, 3$; $\eta_{\mu \nu} = {\rm diag}(1,-1,-1,-1)$ is the Minkowski metric tensor and $z$ is the holographical coordinate running from zero to $\infty$. 

The relevant terms in the AdS/QCD action which generate the nucleon form factors are~\cite{Abidin:2009hr}: 
\eq 
S &=& \int d^4x \, dz \, \sqrt{g} \, e^{-\Phi(z)} \, \Big( \bar\Psi \, e_A^M \, \Gamma^A \, V_M \, \Psi + \frac{i}{2} \, \eta_{S,V} \, \bar\Psi \, e_A^M \, e_B^N \, [\Gamma^A, \Gamma^B] \, F^{(S,V)}_{MN} \, \Psi \, \Big) \,, 
\en 
where the basic blocks of the AdS/QCD model are defined as~\cite{Abidin:2009hr}: $g = |{\rm det} \, g_{MN}|$; $\Psi$ and $V_M$ are the 5D Dirac and vector fields dual to the nucleon and electromagnetic fields, respectively; $F_{MN} = \partial_M V_N - \partial_N V_M$;  
$\Gamma^A = (\gamma^\mu, - i \gamma^5)$; $e_A^M = z \delta_A^M$ is the inverse vielbein; $\eta_{S,V}$ are the couplings constrained by the  anomalous magnetic moment of the nucleon: $\eta_p = (\eta_S + \eta_V)/2 = \kappa \,k_p/(2m_N\sqrt{2})$ and $\eta_n = (\eta_S - \eta_V)/2 = \kappa \,k_n/(2m_N\sqrt{2})$. Here the indices $S,V$ denote isoscalar and isovector contributions to the electromagnetic form factors. 

Finally, the results for the nucleon form factors in AdS/QCD are given by~\cite{Abidin:2009hr}: 
\eq 
F_{1}^p(Q^2) &=& C_{1}(Q^2) + \eta_{p} C_{2}(Q^2)\,,\nonumber\\
F_{2}^p(Q^2) &=& \eta_{p} C_{3}(Q^2)\,,\nonumber\\
F_{1}^n(Q^2) &=& \eta_{n} C_{2}(Q^2)\,,\nonumber\\
F_{2}^n(Q^2) &=& \eta_{n} C_{3}(Q^2), \nonumber 
\en 
where $Q^{2} = - t$ and $C_i(Q^2)$ are the structure integrals: 
\eq 
C_{1}(Q^2) &=& \int dz e^{-\Phi} 
\frac{V(Q,z)}{2 z^{3}} (\psi_{L}^{2}(z) + \psi_{R}^{2}(z))\,, 
\nonumber\\ 
C_{2}(Q^2) &=& \int dz e^{-\Phi} \frac{\partial_{z} V(Q,z)}{2 z^{2}} 
(\psi_{L}^{2}(z) - \psi_{R}^{2}(z))\,,  \label{Ci}\\ 
C_{3}(Q^2) &=& \int dz e^{-\Phi} 
\frac{2 m_{N} V(Q,z)}{2 z^{2}} \psi_{L}(z) \psi_{R}(z)\,. 
\nonumber 
\en 
$\psi_{L}(z)$ and $\psi_{R}(z)$ are the Kaluza-Klein modes (normalizable wave functions), which are dual to left- and right-handed nucleon fields:  
\eq \label{PsiLR}
\psi_L(z) = \kappa^3 z^4\,, \quad 
\psi_R(z) = \kappa^2 z^3 \sqrt{2} 
\en 
and 
\eq\label{V}
V (Q,z) = \Gamma(1 + \frac{Q^{2}}{4 \kappa^{2}}) 
U(\frac{Q^{2}}{4 \kappa^{2}}, 0, \kappa^2 z^2)
\en 
is the bulk-to-boundary propagator of the vector field in the axial gauge.

Expressions for the GPDs in terms of the AdS modes can be obtained using the procedure of light-front mapping suggested by Brodsky and T\'eramond~\cite{Brodsky:2003px,Brodsky:2006uqa,Brodsky:2007hb,Brodsky:2008pf} and extended in~\cite{Vega:2009zb,Vega:2010yq,Branz:2010ub}. In the present case this procedure is based on the use of the integral representation for the bulk-to-boundary propagator introduced by Grigoryan and Radyushkin~\cite{Grigoryan:2007my}: 
\eq 
\label{VInt}
V (Q,z) = \kappa^{2} z^{2} \int_{0}^{1} \frac{dx}{(1-x)^{2}} 
x^{\frac{Q^{2}}{4 \kappa^{2}}} e^{- \frac{x}{1-x} \kappa^{2} z^{2}}\,,   
\en 
where the variable $x$ is equivalent to the light-cone momentum fraction. Matching the respective expressions for the nucleon form factors results (after performing integration over the holographic coordinate $z$) in the nonforward parton densities of the nucleon as: 
\eq 
H_{v}^{q}(x,Q^2) &=& q(x) \, x^a \,, 
\label{HqAdS}\\ 
E_{v}^{q}(x,Q^2) &=& e^q(x) \, x^a \,,
\label{EqAdS}  
\en 
where $a = Q^2/(4\kappa^2)$, and $q(x)$ and $e^q(x)$ are distribution functions given by: 
\eq 
q(x)   = \alpha^q \gamma_{1}(x) + \beta^q \gamma_{2}(x)\,, \quad 
e^q(x) = \beta^q \gamma_{3}(x)\,, 
\en 
where the flavor couplings $\alpha_q, \beta_q$ and functions $\gamma_i(x)$ are written as
\eq 
\alpha^u = 2\,, \ \alpha^d = 1\,, \ 
\beta^u = 2 \eta_{p} + \eta_{n} \,, \ 
\beta^d = \eta_{p} + 2 \eta_{n} \,  
\en 
and 
\eq
\gamma_{1}(x) &=& 
\frac{1}{2} (5 - 8x + 3x^{2})\,, \nonumber\\
\gamma_{2}(x) &=& 1 - 10x + 21x^{2} - 12x^{3} \,, 
\label{gamma} \\
\gamma_{3}(x) &=& 
\frac{6 m_N \sqrt{2}}{\kappa} (1 - x)^{2} \,. \nonumber
\en 
Eqs.~(\ref{HqAdS})-(\ref{gamma}), which display the nonforward parton densities of the nucleon, are the main result of this matching procedure. 
Notice that these functions have an exponential form, which is typical when choosing an ansatz for these functions, and we can also see that they are consistent with a linear Regge behavior at small $x$~\cite{Goeke:2001tz,Guidal:2004nd}. In Figs.1 and 2 we show the nonforward parton distributions $H_{v}^{q}$ and $E_{v}^{q}$ for nucleons, obtained from the expressions deduced on the AdS side according to the holographical model considered in~\cite{Abidin:2009hr}.

The parameters involved are the same as used by Abidin and Carlson~\cite{Abidin:2009hr}, i.e. $\kappa = 350$ MeV, $\eta_{p} = 0.224$, $\eta_{n} = -0.239$, which were fixed in order to reproduce the mass $m_N = 2\kappa \sqrt{2}$ and the anomalous magnetic moments of the nucleon. 

\begin{center}
\begin{figure*}[ht]
  \begin{tabular}{c c}
    \includegraphics[width=3.0 in]{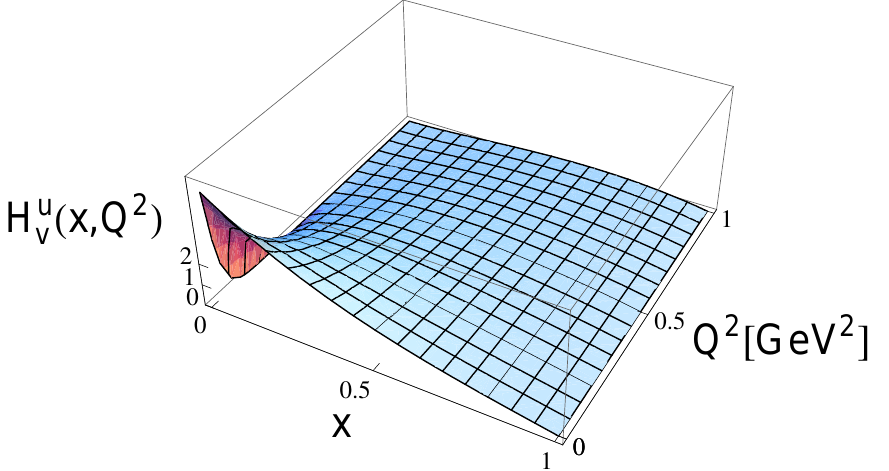}
    \includegraphics[width=3.0 in]{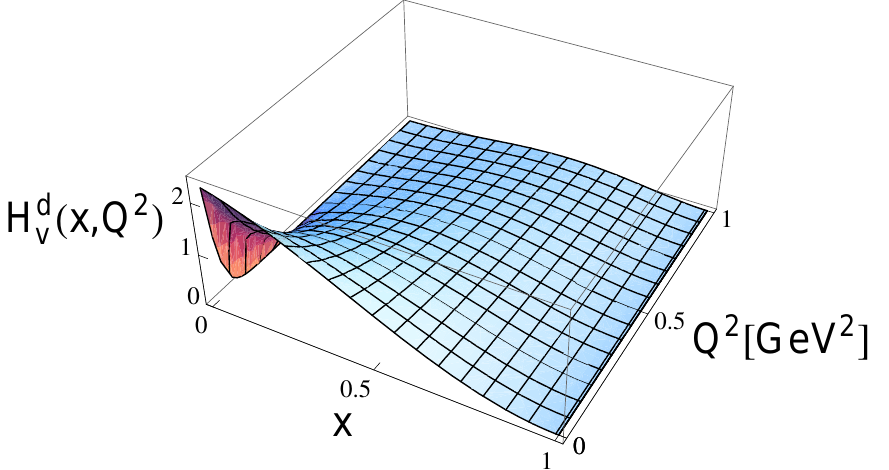} \\
    \includegraphics[width=3.0 in]{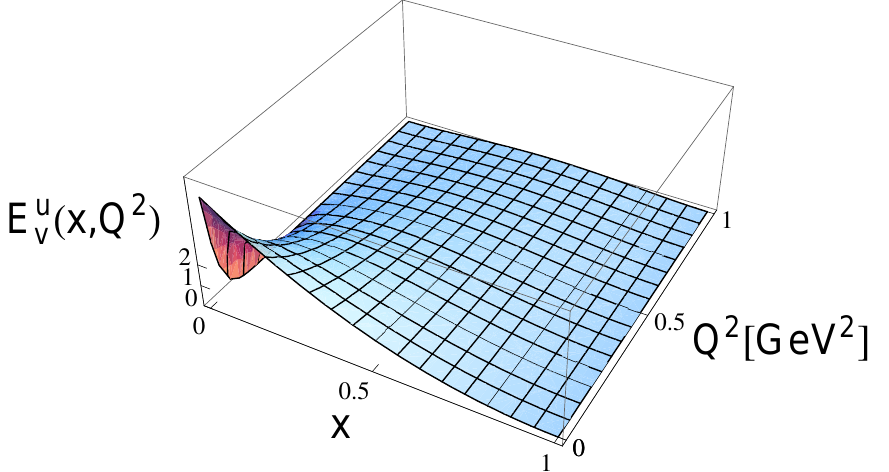}
    \includegraphics[width=3.0 in]{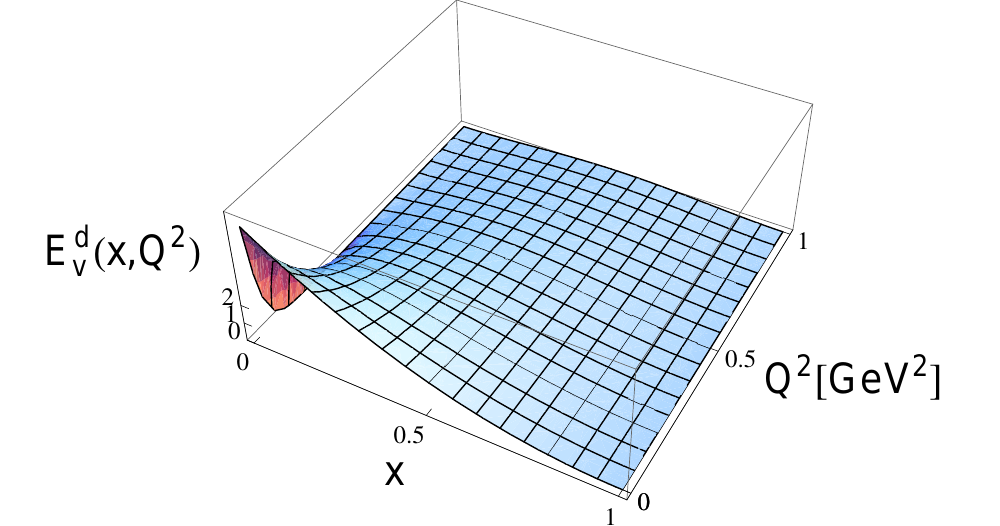}
  \end{tabular}
\caption{The helicity-independent generalized parton distributions (GPDs) of quarks for the nucleon in the zero skewness calculated in the holographical model here described.}
%\caption{$q(x,\bfb)$ plots. The upper corespond to $u(x,\bfb)$ and the lower to $d(x,\bfb)$. Both cases consider $x = 0.1$.}
\end{figure*}
\end{center}

\subsection{Nucleon GPDs in impact space}

Another interesting aspect to consider is the nucleon GPDs in impact space. As shown by Burkardt~\cite{Burkardt:2000za}, the GPDs in momentum space are related to impact parameter dependent parton disrtibutions via Fourier transform. GPDs in impact space give access to the distribution of partons in the transverse plane, which is quite important for understanding the nucleon structure. Here we consider just a couple of quantities in impact space.

\begin{center}
\begin{figure*}[t]
  \begin{tabular}{c c}
    \includegraphics[width=3.0 in]{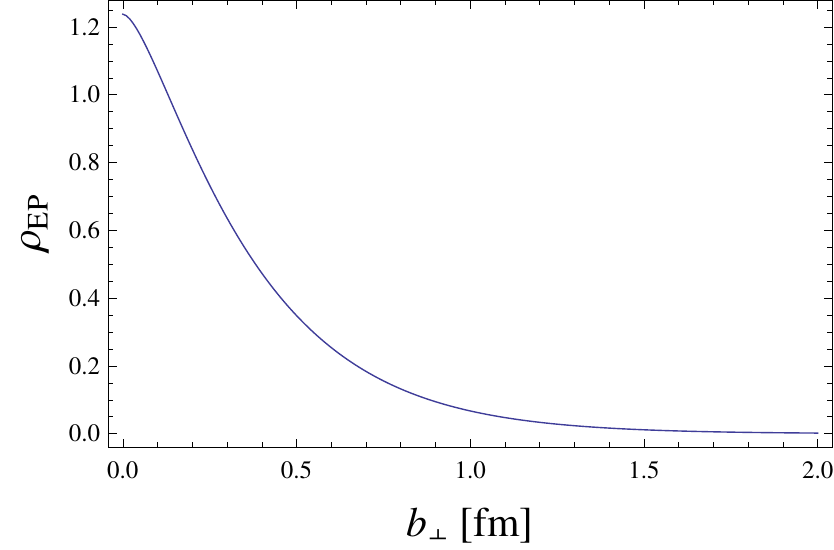}
    \includegraphics[width=3.0 in]{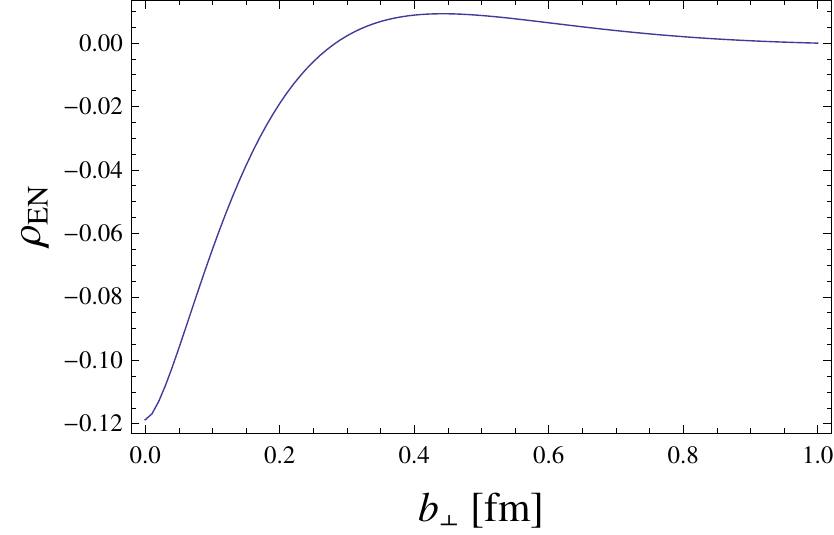} \\
    \includegraphics[width=3.0 in]{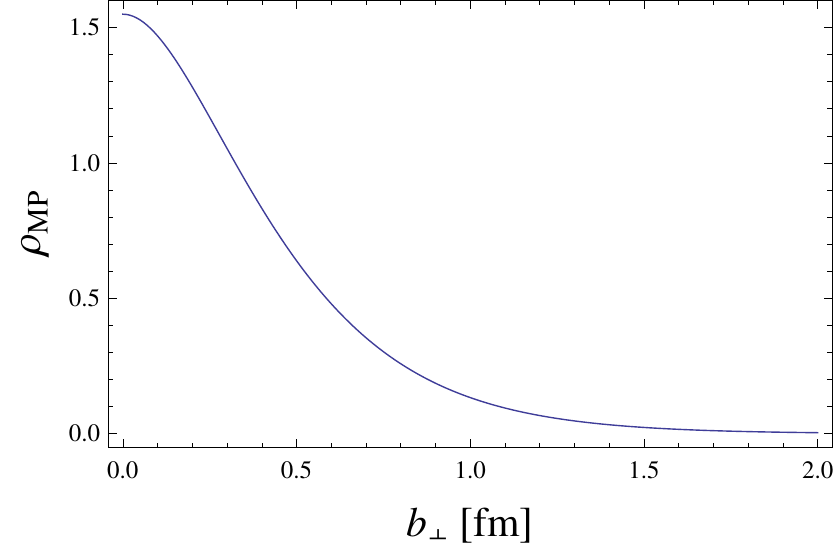}
    \includegraphics[width=3.0 in]{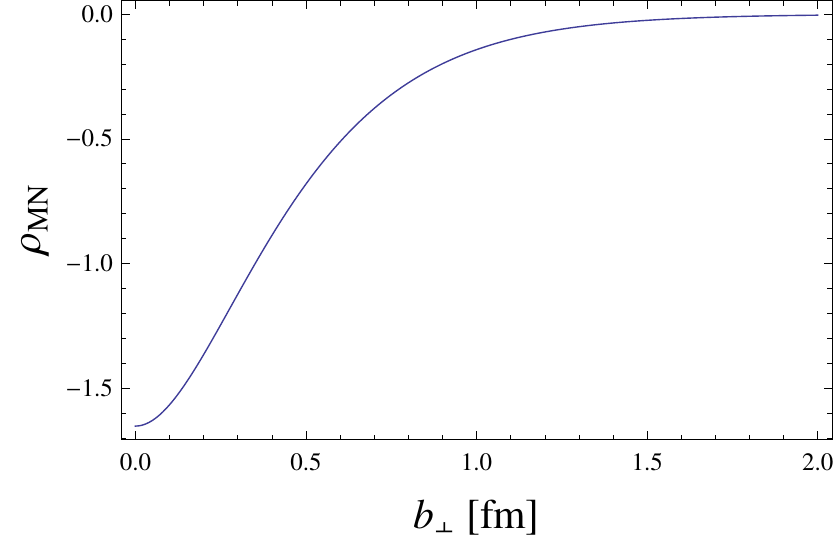}
  \end{tabular}
\caption{Parton charge $(\rho_E(b))$ and magnetization $(\rho_M(b))$ densities in the transverse impact space.}
\end{figure*}
\end{center}

Following Refs.~\cite{Burkardt:2000za} and~\cite{Diehl:2004cx,Miller:2007uy} we define the following set of nucleon quantities in impact space. The nucleon GPDs in impact space 
\eq 
q(x,b) &=& \int\frac{d^2 k}{(2\pi)^2} H_q(x,k^2) 
e^{-i b k}\,, \nonumber\\
e^q(x,b) &=& \int\frac{d^2 k}{(2\pi)^2} E_q(x,k^2) 
e^{-i b k}\,. 
\en 
Parton charge $(\rho_E(b))$ and magnetization $(\rho_M(b))$ densities in transverse impact space 
\eq 
\rho_E(b) &=& \sum\limits_{q} e_q \int\limits_0^1 dx q(x,b) \,, 
\nonumber\\
\rho_M(b) &=& \sum\limits_{q} e_q \int\limits_0^1 dx e^q(x,b) \,, 
\en 

\section{Conclusions} 

We discuss an alternative to calculate the nucleon GPDs in both momentum and impact space using ideas of AdS/QCD. LFH and sum rules relating electromagnetic form factors to the GPDs functions $H_{v}^{q}(x,Q^2)$ and $E_{v}^{q}(x,Q^2)$. The procedure used is similar to the one considered in some applications of LFH, where by comparing form factors it is possible to obtain mesonic light front wave functions. In the present case it is not necessary to reinterpretate the holographical coordinate $z$ as in standard LFH, where $z$ is the distance between constituent partons. 

The nucleon GPDs obtained have an exponential form as in several phenomenological approaches, and their detailed form is typically used in limit the $x \to 0$.

\begin{acknowledgements}
The authors thank Stan Brodsky and Guy de T\'eramond for useful discussions and remarks. V.E.L. would like to thank Departamento de F\'\i sica y Centro Cient\'\i fico Tecnol\'ogico de Valpara\'\i so (CCTVal), Universidad T\'ecnica Federico Santa Mar\'\i a, Valpara\'\i so, Chile for warm
hospitality.
\end{acknowledgements}

\end{document}